\newif\ifFull
\Fullfalse

\documentclass{llncs}

\usepackage{amsmath}
\usepackage{graphicx}
\usepackage{xcolor}
\usepackage{hyperref}
\usepackage{algorithmic}
\usepackage{multirow}

\usepackage{times}
\usepackage{mathptmx}

\newcommand{\abs}[1]{{\left\vert #1 \right\vert}}

\title{Listing All Maximal Cliques in\\ Large Sparse Real-World Graphs}

\author{David Eppstein \and Darren Strash}

\institute{Department of Computer Science, University of California, Irvine, USA.}

\begin{document}
\maketitle

\begin{abstract}
We implement a new algorithm for listing all maximal cliques in sparse graphs due to Eppstein, L\"offler,
and Strash (ISAAC 2010) and analyze its performance on a large corpus of real-world graphs. Our analysis shows that this
algorithm is the first to offer a practical solution to listing all maximal cliques
in large sparse graphs. All other theoretically-fast algorithms for sparse graphs
have been shown to be significantly slower than the algorithm of Tomita et al.
(Theoretical Computer Science, 2006) in practice. However, the algorithm
of Tomita et al. uses an adjacency matrix, which requires too much space
for large sparse graphs. Our new algorithm opens the door for fast 
analysis of large sparse graphs whose adjacency matrix will not fit into 
working memory.


\keywords{maximal clique listing, Bron--Kerbosch algorithm, sparse graphs, $d$-degenerate graphs}
\end{abstract}

\section{Introduction}
Clique finding procedures arise in the solutions to a wide variety of important application problems.
The problem of finding cliques was first studied in social network analysis, as a way of finding closely-interacting communities of agents in a social network~\cite{harary57}.
In bioinformatics, clique finding procedures have been used to find frequently occurring patterns in protein structures~\cite{grindley93,koch2001,KocLenWan-JCB-96}, to predict the structures of proteins from their molecular sequences~\cite{samudrala98}, and to find similarities in shapes that may indicate functional relationships between proteins~\cite{gardiner99}. Other applications of clique finding problems include information retrieval~\cite{auguston70}, computer vision~\cite{HorSko-PAMI-89}, computational topology~\cite{Zom-SoCG-10}, and e-commerce~\cite{ZakParOgi-KDD-97}.

For many applications, we do not want to report one large clique, but all 
maximal cliques. Any algorithm which solves this problem must take
exponential time in the worst-case because graphs can contain an
exponential number of cliques~\cite{moon-moser65}. However,
graphs with this worst-case behavior are not typically encountered in practice. More than likely,
the types of graphs that we will encounter are sparse~\cite{goel2006}.  Therefore,
the feasibility of clique listing algorithms lies in their ability to appropriately
handle sparse input graphs. Indeed, it has long been known that certain 
sparse graph families, such as planar graphs and graphs with low arboricity, contain
only a linear number of cliques, and that all maximal cliques in these
graphs can be listed in linear time~\cite{chiba85,chrobak91}.
In addition, there are also several methods to list all cliques in time
polynomial in the number of cliques reported~\cite{tsukiyama77},
which can be done faster if parameterized on a sparsity measure such
as maximum degree~\cite{makino2004}.

Many different clique-finding algorithms have been implemented, and an
algorithm of Tomita et al.~\cite{tomita2006}, based on the much earlier Bron--Kerbosch algorithm~\cite{bron73}, has been shown 
through many experiments to be faster by orders of magnitude
in practice than others. An unfortunate drawback
of the algorithm of Tomita et al., however, is that both its theoretical analysis and implementation
rely on the use of an adjacency matrix representation of
the input graph. For this reason, their algorithm has
limited applicability for large sparse graphs, whose
adjacency matrix may not fit into working memory. We therefore seek to have the best of both worlds: we would
ideally like an algorithm that rivals the speed of 
the Tomita et al. result, while having linear storage cost. 

Recently, together with Maarten L\"offler, the authors developed and published a new algorithm for listing maximal cliques, particularly optimized for the case that the input graph is sparse~\cite{els2010}. This new algorithm combines features of both the algorithm of Tomita et al. and the earlier Bron--Kerbosch algorithm on which it was based,  and maintains through its recursive calls a dynamic graph data structure representing the adjacencies between the vertices that remain relevant within each call. When analyzed using parameterized complexity in terms of the degeneracy of the input graph (a measure of its sparsity) its running time is near-optimal in terms of the worst-case number of cliques that a graph with the same sparsity could have. However, the previous work of the authors with L\"offler did not include any implementation or experimental results showing the algorithm to be good in practice as well as in theory.

\subsection{Our Results}
We implement the algorithm of Eppstein, L\"offler, and Strash for listing all
maximal cliques in sparse graphs~\cite{els2010}.
Using a corpus of many large real-world graphs, together with synthetic data including the Moon--Moser graphs as well as random graphs, we compare the performance of our implementation with the algorithm of Tomita et al. We also implement for comparison, a modified version of the Tomita et al. algorithm that uses adjacency lists in place of adjacency matrices, and a simplified version of the Eppstein--L\"offler--Strash algorithm that represents its subproblems as lists of vertices instead of as dynamic graphs. Our results show that, for large sparse graphs, the new algorithm is as fast or faster than Tomita et al., and sometimes faster by very large factors. For graphs that are not as sparse, the new algorithm is sometimes slower than the algorithm of Tomita et al., but remains within a small constant factor of its performance.

\section{Preliminaries}
We work with an undirected graph $G = (V,E)$ with $n$ vertices
and $m$ edges. For a vertex $v$, let $\Gamma(v)$ be its neighborhood $\{w\mid (v,w)\in E\}$, and similarly for a subset $W \subset V$ let $\Gamma(W)$ be the set $\bigcap_{w \in W} \Gamma(w)$, the common neighborhood of all vertices in $W$.

\subsection{Degeneracy}

\begin{definition}[degeneracy]
The degeneracy of a graph $G$ is the smallest number $d$
such that every subgraph of $G$ contains a vertex of degree at most $d$. 
\end{definition}

Every graph with degeneracy $d$ has a \emph{degeneracy ordering}, a linear ordering of 
the vertices such that each vertex has at most $d$ 
neighbors later than it in the ordering. The degeneracy of a given graph and a degeneracy ordering of the graph can both be computed in linear time~\cite{batagelj2003}.

\subsection{The Algorithm of Tomita et al.}

\begin{figure}[t]
{\bf proc} Tomita($P$, $R$, $X$)
\begin{algorithmic}[1]
\IF{$P\cup X = \emptyset$}
    \STATE report $R$ as a maximal clique
\ENDIF
\STATE choose a pivot $u \in P\cup X$ to maximize $\abs{P\cap\Gamma(u)}$
\FOR{ {\bf each} vertex $v\in P\setminus \Gamma(u)$}
    \STATE Tomita($P\cap \Gamma(v)$, $R\cup\{v\}$, $X\cap \Gamma(v)$)
    \STATE $P \leftarrow P \setminus \{v\}$
    \STATE $X \leftarrow X \cup \{v\}$
\ENDFOR
\end{algorithmic}

\caption{The Bron--Kerbosch algorithm with the pivoting strategy of Tomita et al.}
\label{figure:bkalg}
\end{figure}

The algorithm of Tomita et al.~\cite{tomita2006} is an implementation of 
Bron and Kerbosch's algorithm~\cite{bron73}, using a heuristic called
\emph{pivoting}~\cite{koch2001,cazals2008}. The Bron--Kerbosch algorithm is a simple recursive
algorithm that maintains three sets of vertices: a partial clique $R$, 
a set of candidates for clique expansion $P$, and a set of forbidden vertices $X$.
In each recursive call, a vertex $v$ from $P$ is added to the partial
clique $R$, and the sets of candidates
for expansion and forbidden vertices are restricted to include only neighbors of~$v$.
If $P\cup X$ becomes empty, the algorithm reports $R$ as a maximal clique, but if $P$ becomes empty while $X$ is nonempty, the algorithm backtracks without reporting a clique.

In the basic version of the algorithm, $|P|$ recursive calls are made, one for each vertex in $P$. The pivoting heuristic reduces the number of recursive calls by choosing a vertex $u$ in $P \cup X$ called a \emph{pivot}.
All maximal cliques must contain a non-neighbor of $u$ (counting $u$ itself as a non-neighbor), and therefore, the recursive calls can be restricted to the intersection of $P$ with the non-neighbors.

The algorithm of Tomita et al. chooses the pivot so that $u$ has the maximum number of neighbors in $P$, and therefore the minimum number of non-neighbors, among all possible pivots. Computing both the pivot and the vertex sets for the recursive calls
can be done in time $O(\abs{P}\cdot(\abs{P} + \abs{X}))$ within each call to the algorithm, using an
adjacency matrix to quickly test the adjacency of pairs of vertices.  This pivoting strategy, together with this adjacency-matrix-based method for computing the pivots, leads to a worst-case time bound of $O(3^{n/3})$ for listing all maximal cliques~\cite{tomita2006}.

\subsection{The Algorithm of Eppstein, L\"offler, and Strash}

\begin{figure}[t]
{\bf proc} Degeneracy($V$, $E$)
\begin{algorithmic}[1]
\FOR{each vertex $v_i$ in a degeneracy ordering $v_0$, $v_1$, $v_2$, \dots of $(V,E)$}
    \STATE{$P\leftarrow \Gamma(v_i)\cap \{v_{i+1},\ldots, v_{n-1}\}$ }
    \STATE{$X\leftarrow \Gamma(v_i)\cap \{v_0,\ldots, v_{i-1}\}$ }
    \STATE{Tomita($P$, $\{v_i\}$, $X$)}
\ENDFOR
\end{algorithmic}

\caption{The algorithm of Eppstein, L\"offler, and Strash.}
\end{figure}

Eppstein, L\"offler, and Strash~\cite{els2010} provide a different variant of the Bron--Kerbosch algorithm that obtains near-optimal worst-case time bounds for
graphs with low degeneracy. They first compute a degeneracy ordering of the graph; the outermost call in the recursive algorithm selects the vertices $v$ to be used in each recursive call, in this order, without pivoting.
Then for each vertex $v$ in the order, a call is made to the algorithm of Tomita et al.~\cite{tomita2006} to compute all cliques containing
$v$ and $v$'s later neighbors, while avoiding $v$'s earlier neighbors.
The degeneracy ordering limits the size of $P$ within these recursive calls to be at most $d$, the degeneracy
of the graph.

A simple strategy for determining the pivots in each call to the algorithm of Tomita et al., used as a subroutine within this algorithm, would be to loop over all possible pivots in $X\cup P$ and, for each one, loop over its later neighbors in the degeneracy ordering to determine how many of them are in $P$. The same strategy can also be used to perform
the neighbor intersection required for recursive calls. With the pivot selection and set intersection algorithms implemented in this way, the algorithm would have running time $O(d^2n3^{d/3})$, a factor of $d$ larger than the worst-case output size, which is $O(d(n-d)3^{n/3})$.

However, Eppstein et al. provide a refinement of this algorithm that stores, at each level of the recursion, the subgraph of $G$ with vertices in $P\cup X$ and edges having at least one endpoint in $P$. Using this subgraph, they reduce the pivot computation
time to $\abs{P}(\abs{X} + \abs{P})$, and the neighborhood intersection
for each recursive call to time $\abs{P}^2(\abs{X} + \abs{P})$, which reduces
the total running time to $O(dn3^{d/3})$. This running time matches the worst-case output size of
the problem whenever $d\le n-\Omega(n)$.
As described by Eppstein et al., storing the subgraphs at each level of the recursion may require as much as $O(dm)$ space.  But as we show in Section~\ref{sec:details}, it is possible to achieve the same optimal running time
with space overhead $O(n+m)$.

\subsection{Tomita et al. with Adjacency Lists}
In our experiments, we were only able to run the algorithm of Tomita et al.~\cite{tomita2006} on graphs of small to moderate size, due to its use of the adjacency matrix representation. In order to have a basis for comparison with this algorithm on larger graphs, we also implemented a simple variant of the algorithm which stores the input graph in an adjacency list representation, and
which performs the pivot computation by iterating over all vertices in $P\cup X$ and testing all
neighbors for membership in $P$. When a vertex $v$ is added to $R$ for a recursive call,
we can intersect the neighborhood of $r$ with $P$ and $X$ by iterating over its neighbors in the same way.

Let $\Delta$ be the maximum degree of the given input graph; then the pivot computation takes time $(O\Delta(\abs{X} + \abs{P}))$.
Additionally, preparing subsets for all recursive calls takes time $O(\abs{P} \Delta)$. Fitting these facts into the analysis
of Tomita et al. gives us a $O(\Delta(n-\Delta)3^{\Delta/3})$ time algorithm.  $\Delta$ may be significantly larger than the degeneracy, so this algorithm's theoretical time bounds are not as good as those of Tomita et al. or Eppstein et al.; nevertheless, the simplicity of this algorithm makes it competitive with the others for many problem instances.

\section{Implementation and experiments}
We implemented the algorithm of Tomita et al. using the adjacency matrix representation,
and the simple adjacency list representation for comparison. 
We also implemented three variants of the algorithm of Eppstein, L\"offler, and Strash:
one with no data structuring, using the fact that vertices have few later neighbors
in the degeneracy ordering, an implementation of the dynamic graph data structure that
only uses $O(m+n)$ extra space total, and an alternative implementation of the data structure
based on bit vectors. The bit vector implementation
executed no faster than the data structure implementation, so we omit its experimental timings and any discussion of its implementation details. 
\subsection{Implementation Details}
\label{sec:details}

\begin{figure}[b]
\begin{center}
\includegraphics[scale=0.8]{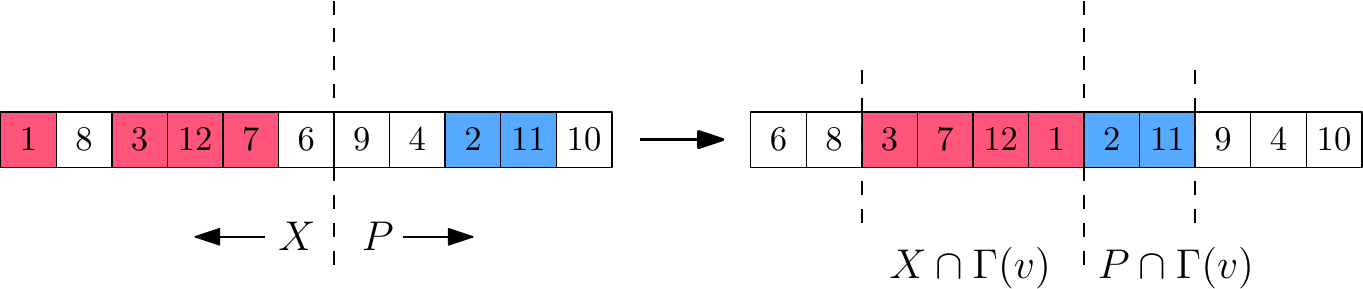}
\caption{When a vertex $v$ is added to the partial clique $R$, its neighbors in $P$ and $X$ (highlighted in this example) are moved toward the dividing line in preparation for the next recursive call. }
\label{fig:pandx}
\end{center}
\end{figure}

\begin{figure}[t]
\begin{center}
\includegraphics[scale=0.8]{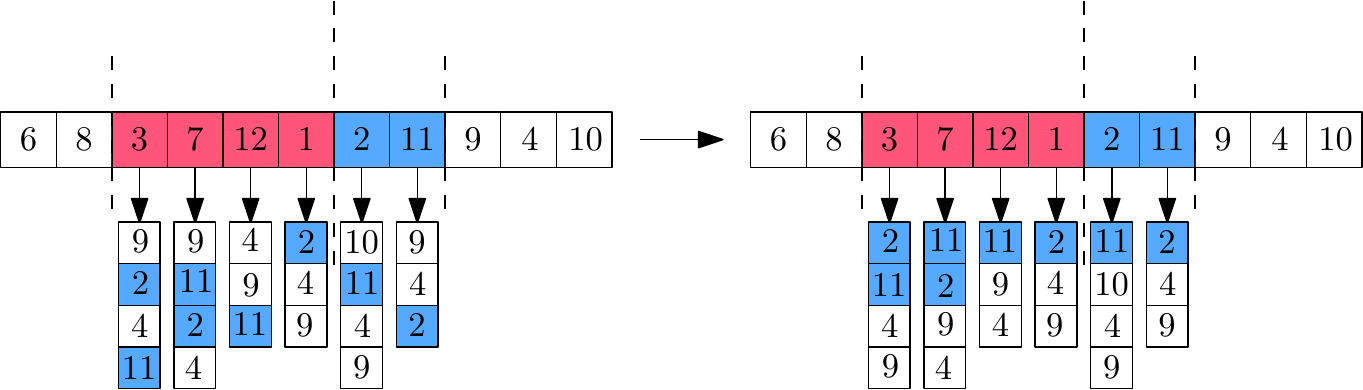}
\caption{For each vertex in $P\cup X$, we keep an array containing neighbors in $P$. We update
these arrays whenever a vertex is moved from $P$ to $R$, and whenever we need to intersect
a neighborhood with $P$ and $X$ for a recursive call.}
\label{fig:pandx-ds}
\end{center}
\end{figure}

We maintain the sets of vertices $P$ and $X$ in a single array, which is
passed between recursive calls. Initially, the array contains the elements of $X$
followed by the elements of $P$. 
We keep a reverse lookup table, so that we can look up the
index of a vertex in constant time. With this lookup table,
we can tell whether a vertex is in $P$ or $X$ in constant time,
by testing that its index is in the appropriate subarray.
When a vertex $v$ is added to $R$ in preparation
for a recursive call, we reorder the array. Vertices in $\Gamma(v)\cap X$ are moved to the 
end of the $X$ subarray, and vertices in $\Gamma(v)\cap P$ are moved
to the beginning of the $P$ subarray (see Figure~\ref{fig:pandx}). We then make a recursive call
on the subarray containing the vertices $\Gamma(v)\cap (X\cup P)$. After the recursive call, we move $v$ to $X$ by swapping it to the beginning of the $P$ subarray
and moving the boundary so that $v$ is in the $X$ subarray.
Of course, moving vertices between sets will affect $P$ and $X$ in higher recursive calls.
Therefore, in a given recursive call, we maintain a list of the vertices that are 
moved from $P$ to $X$, and move these vertices back to $P$ when the call ends.

The pivot computation data structure is stored as set of arrays, one for 
each potential pivot vertex $u$ in $P\cup X$, containing the neighbors of $u$ in $P$.
Whenever $P$ changes, we reorder the elements in these arrays so that neighbors in $P$ are stored first (see Figure~\ref{fig:pandx-ds}). Computing the pivot is as simple as iterating through each array until we encounter a neighbor that is not in $P$.
This reordering procedure allows us to maintain one set of arrays throughout 
all recursive calls, requiring linear space total.  Making a new 
copy of this data structure for each recursive call would require 
space $O(dm)$.

\subsection{Results}

We implemented all algorithms in the C programming language, and
ran experiments on a Linux workstation running the 32-bit version of
Ubuntu 10.10, with a 2.53 GHz Intel Core i5 M460 processor (with three cache levels 
of 128KB, 512KB, and 3,072KB respectively) and 2.6GB of memory. We compiled
our code with version 4.4.5 of the gcc compiler with the \texttt{-O2} optimization 
flag.

In our tables of results, ``tomita'' is the algorithm of Tomita et al.,
``maxdegree'' is the simple implementation of Tomita et al.'s algorithm
for adjacency lists, and ``hybrid'' and ``degen'' are the implementations
of Eppstein, L\"offler, and Strash with no data structure and with the
linear space data structure, respectively. We provide the elapsed running running times (in seconds) for each of these algorithms; an asterisk indicates that the algorithm was unable to run on that problem instance due to time or space limitations.
In addition, we list the number of vertices $n$, edges $m$, the degeneracy $d$,
and the number of maximal cliques $\mu$.

Our primary experimental data consisted of four publicly-available databases of real-world networks, including non-electronic and electronic social networks as well as networks from bioinformatics applications.
\begin{itemize}
\item A data base curated by Mark Newman~\cite{marknewman} (Table~\ref{table-uci}) which consists primarily of social networks; it also includes word co-occurrence data and a biological neural network. Many of its graphs were too small for us to  time our algorithms accurately, but our algorithm was faster than that of Tomita et al. on all four of the largest graphs; in one case it was faster by a factor of approximately 130. 
\item The BioGRID data~\cite{BioGRID2006} (Table~\ref{table-biogrid}) consists of several protein-protein interaction networks with from one to several thousand vertices, and varying sparsities. Our algorithm was significantly faster than that of Tomita et al. on the worm and fruitfly networks, and matched or came close to its performance on all the other networks, even the relatively dense yeast network. 
\item We also tested six large social and bibliographic networks that appeared in the Pajek data set but were not in the other data sets~\cite{pajek} (Table~\ref{table-pajek}). Our algorithm was consistently faster on these networks. Due to their large size, the algorithm of Tomita et al. was unable to run on two of these networks; nevertheless, our algorithm found all cliques quickly in these graphs.
\item We also tested a representative sample of graphs from the Stanford Large Network Dataset Collection~\cite{snap} (Table~\ref{table-snap}). These included road networks, a co-purchasing network from Amazon.com data,
social networks, email networks, a citation network, and two Web graphs. Nearly all of these input graphs were
too large for the Tomita et al. algorithm to fit into memory. For graphs which are extremely sparse, it
is no surprise that the maxdegree algorithm was faster than our algorithm, but our algorithm was consistently
fast on each of these data sets, whereas the maxdegree algorithm was orders of magnitude slower than our algorithm on
the large soc-wiki-Talk network.

\end{itemize}

\begin{table}[htb!]
\begin{center}
\caption{Experimental results for Mark Newman's data sets~\cite{marknewman}}
\label{table-uci}
\begin{tabular}{l@{\hspace{0.2cm}}r@{\hspace{0.2cm}}r@{\hspace{0.2cm}}r@{\hspace{0.2cm}}r@{\hspace{0.2cm}}r@{\hspace{0.2cm}}r@{\hspace{0.2cm}}r@{\hspace{0.2cm}}r@{\hspace{0.2cm}}r}
graph                                       &n     &m       &d  &$\mu$     &tomita     &maxdegree&hybrid   & degen   \\ \hline
zachary~\cite{zachary}                      &34    &78      &4  &39        &$< 0.01$   &$< 0.01$ &$< 0.01$ &$< 0.01$ \\
dolphins~\cite{dolphins}                    &62    &159     &4  &84        &$< 0.01$   &$< 0.01$ &$< 0.01$ &$< 0.01$ \\
power~\cite{power-celegensneural}           &4,941 &6,594   &5  &5,687     &0.29       &$< 0.01$ &0.01     &0.01     \\
polbooks~\cite{polbooks}                    &105   &441     &6  &199       &$< 0.01$   &$< 0.01$ &$< 0.01$ &$< 0.01$ \\
adjnoun~\cite{adjnoun}                      &112   &425     &6  &303       &$< 0.01$   &$< 0.01$ &$< 0.01$ &$< 0.01$ \\
football~\cite{football}                    &115   &613     &8  &281       &$< 0.01$   &$< 0.01$ &$< 0.01$ &$< 0.01$ \\
lesmis~\cite{lesmis}                        &77    &254     &9  &59        &$< 0.01$   &$< 0.01$ &$< 0.01$ &$< 0.01$ \\
celegensneural~\cite{power-celegensneural}  &297   &1,248   &9  &856       &$< 0.01$   &$< 0.01$ &$< 0.01$ &$< 0.01$ \\
netscience~\cite{netscience}                &1,589 &2,742   &19 &741       &0.02       &$< 0.01$ &$< 0.01$ &$< 0.01$ \\
internet~\cite{marknewman}                  &22,963&48,421  &25 &39,275    &6.68       &0.28     &0.11     &0.11     \\
condmat-2005~\cite{condmat-astro}           &40,421&175,693 &29 &34,274    &39.65      &0.22     &0.32     &0.35     \\
polblogs~\cite{polblogs}                    &1,490 &16,715  &36 &49,884    &0.08       &0.28     &0.18     &0.12     \\
astro-ph~\cite{condmat-astro}               &16,706&121,251 &56 &15,794    &3.44       &0.19     &0.22     &0.23     
\end{tabular}
\end{center}
\end{table}

\begin{table}[htb!]
\begin{center}
\caption{Experimental results for BioGRID data sets (PPI Networks)\cite{BioGRID2006}.}
\label{table-biogrid}
\begin{tabular}{l@{\hspace{0.2cm}}r@{\hspace{0.2cm}}r@{\hspace{0.2cm}}r@{\hspace{0.2cm}}r@{\hspace{0.2cm}}r@{\hspace{0.2cm}}r@{\hspace{0.2cm}}r@{\hspace{0.2cm}}r@{\hspace{0.2cm}}r}
graph        &n     &m      &d    &$\mu$      &tomita         &maxdegree  &hybrid     &degen      \\ \hline
mouse        &1,455 &1,636  &6    &1,523      &0.01           &$< 0.01$   &$< 0.01$   &$< 0.01$   \\
worm         &3,518 &3,518  &10   &5,652      &0.14           &0.01       &0.01       &0.01       \\
plant        &1,745 &3,098  &12   &2,302      &0.02           &$< 0.01$   &$< 0.01$   &$< 0.01$   \\
fruitfly     &7,282 &24,894 &12   &21,995     &0.62           &0.03       &0.03       &0.04       \\
human        &9,527 &31,182 &12   &23,863     &1.06           &0.03       &0.05       &0.05       \\
fission-yeast&2,031 &12,637 &34   &28,520     &0.06           &0.12       &0.09       &0.07       \\
yeast        &6,008 &156,945&64   &738,613    &1.74           &11.37      &4.22       &2.17       
\end{tabular}
\end{center}
\end{table}

\begin{table}[tbh!]
\begin{center}
\caption{Experimental results for Pajek data sets~\cite{pajek}.}
\label{table-pajek}
\begin{tabular}{l@{\hspace{0.2cm}}r@{\hspace{0.2cm}}r@{\hspace{0.2cm}}r@{\hspace{0.2cm}}r@{\hspace{0.2cm}}r@{\hspace{0.2cm}}r@{\hspace{0.2cm}}r@{\hspace{0.2cm}}r@{\hspace{0.2cm}}r}
graph                     &n      &m        &d   &$\mu$       &tomita        &maxdegree   &hybrid      &degen      \\ \hline
foldoc~\cite{foldoc}      &13,356 &91,471   &12  &39,590      &2.16          &0.11        &0.14        &0.13       \\
eatRS~\cite{eatRS}        &23,219 &304,937  &34  &298,164     &7.62          &1.52        &1.55        &1.03       \\
hep-th~\cite{hep-th}      &27,240 &341,923  &37  &446,852     &12.56         &3.40        &2.40        &1.70       \\
patents~\cite{patents}    &240,547&560,943  &24  &482,538     &*             &0.56        &1.22        &1.65       \\
days-all~\cite{days-all}  &13,308 &148,035  &73  &2,173,772   &5.83          &62.86       &9.94        &5.18       \\
ND-www~\cite{ND-www}      &325,729&1,090,108&155 &495,947     &*             &1.80        &1.81        &2.12       
\end{tabular}
\end{center}
\end{table}

\begin{table}[tbh!]
\begin{center}
\caption{Experimental results for Stanford data sets~\cite{snap}.}
\label{table-snap}
{\small
\begin{tabular}{l@{\hspace{0.1cm}}r@{\hspace{0.1cm}}r@{\hspace{0.1cm}}r@{\hspace{0.1cm}}r@{\hspace{0.1cm}}r@{\hspace{0.1cm}}r@{\hspace{0.1cm}}r@{\hspace{0.1cm}}r@{\hspace{0.1cm}}r}
graph                           &n         &m           &d      &$\mu$       &tomita   &maxdegree   &hybrid      &degen      \\ \hline
roadNet-CA~\cite{slashdot}      &1,965,206 &2,766,607   &3      &2,537,996   &*        &2.00        &5.34        &5.81      \\ 
roadNet-PA~\cite{slashdot}      &1,088,092 &1,541,898   &3      &1,413,391   &*        &1.09        &2.95        &3.21      \\ 
roadNet-TX~\cite{slashdot}      &1,379,917 &1,921,660   &3      &1,763,318   &*        &1.35        &3.72        &4.00      \\ 
amazon0601~\cite{amazon}        &403,394   &2,443,408   &10     &1,023,572   &*        &3.59        &5.01        &6.03       \\
email-EuAll~\cite{emailEU}      &265,214   &364,481     &37     &265,214     &*        &4.93        &1.25        &1.33       \\
email-Enron~\cite{email-Enron}  &36,692    &183,831     &43     &226,859     &31.96    &2.78        &1.30        &0.90      \\ 
web-Google~\cite{web-Google}    &875,713   &4,322,051   &44     &1,417,580   &*        &9.01        &8.43        &9.70       \\
soc-wiki-Vote~\cite{wiki}       &7,115     &100,762     &53     &459,002     &0.96     &4.21        &2.10        &1.14       \\
soc-slashdot0902~\cite{slashdot}&82,168    &504,230     &55     &890,041     &*        &7.81        &4.20        &2.58       \\
cit-Patents~\cite{patents}      &3,774,768 &16,518,947  &64     &14,787,032  &*        &28.56       &49.22       &58.64     \\ 
soc-Epinions1~\cite{epinions}   &75,888    &405,740     &67     &1,775,074   &*        &27.87       &9.24        &4.78       \\
soc-wiki-Talk~\cite{wiki}       &2,394,385 &4,659,565   &131    &86,333,306  &*        &$> 18,000$  &542.28      &216.00     \\
web-berkstan~\cite{slashdot}    &685,231   &6,649,470   &201    &3,405,813   &*        &76.90       &31.81       &20.87     \\ 
\end{tabular}
}
\end{center}
\end{table}

As a reference point, we also ran our experimental comparisons using the two sets of graphs that
Tomita et al. used in their experiments. First, Tomita et al. used a data set from a DIMACS challenge, a collection of graphs that were intended as difficult examples for clique-finding algorithms, and that have been algorithmically generated (Table~\ref{table-dimacs}). And second, they generated graphs randomly with varying edge densities; in order to replicate their results we generated another set of random graphs with the same parameters (Table~\ref{table-random}).
The  algorithm of Eppstein, L\"offler, and Strash runs about 2 to 3 times slower than that of Tomita et al. on many of these graphs;
this confirms that the algorithm is still competitive on graphs that are not sparse,
in contrast to the competitors in Tomita et al.'s paper which ran
10 to 160 times slower on these input graphs. The largest of the random graphs in the second data set were generated with edge probabilities that made them significantly sparser than the rest of the set; for those graphs our algorithm outperformed that of Tomita et al by a factor that was as large as 30 on the sparsest of the graphs. The maxdegree algorithm was even faster than our algorithm in these cases, but it was significantly slower on other data.

\begin{table}[tbh!]
\begin{center}
\caption{Experimental results for Moon--Moser~\cite{moon-moser65} and DIMACS benchmark graphs~\cite{johnson1996}.}
\label{table-dimacs}
\begin{tabular}{l@{\hspace{0.2cm}}r@{\hspace{0.2cm}}r@{\hspace{0.2cm}}r@{\hspace{0.2cm}}r@{\hspace{0.2cm}}r@{\hspace{0.2cm}}r@{\hspace{0.2cm}}r@{\hspace{0.2cm}}r}
Graphs          &n     &m       &d      &$\mu$       &tomita   &maxdegee &hybrid &degen \\ \hline
M-M-30          &30    &405     &27     &59,049      &0.04     &0.04     &0.06   &0.04  \\
M-M-45          &45    &945     &42     &14,348,907  &7.50     &15.11    &20.36  &10.21 \\
M-M-48          &48    &1080    &45     &43,046,721  &22.52    &48.37    &63.07  &30.22 \\
M-M-51          &51    &1224    &48     &129,140,163 &67.28    &150.02   &198.06 &91.80 \\
MANN\_a9        &45    &918     &27     &590,887     &0.44     &0.88     &0.90   &0.53  \\
brock\_200\_2   &200   &9876    &84     &431,586     &0.55     &2.95     &2.61   &1.22  \\
c-fat200-5      &200   &8473    &83     &7           &0.01     &0.01     &0.01   &0.01  \\
c-fat500-10     &500   &46627   &185    &8           &0.04     &0.04     &0.09   &0.12  \\
hamming6-2      &64    &1824    &57     &1,281,402   &1.36     &4.22     &4.15   &2.28  \\
hamming6-4      &64    &704     &22     &464         &$< 0.01$ &$< 0.01$ &$< 0.01$&$< 0.01$  \\
johnson8-4-4    &70    &1855    &53     &114,690     &0.13     &0.35     &0.40   &0.24  \\
johnson16-2-4   &120   &5460    &91     &2,027,025   &5.97     &27.05    &31.04  &12.17 \\
keller4         &171   &9435    &102    &10,284,321  &5.98     &24.97    &26.09  &11.53 \\
p\_hat300-1     &300   &10933   &49     &58,176      &0.07     &0.29     &0.25   &0.15  \\
p\_hat300-2     &300   &21928   &98     &79,917,408  &91.31    &869.34   &371.72 &163.16
\end{tabular}
\end{center}
\end{table}

\begin{table}[htb!]
\begin{center}
\caption{Experimental results on random graphs.}
\label{table-random}
\begin{tabular}{l@{\hspace{0.2cm}}l@{\hspace{0.2cm}}r@{\hspace{0.2cm}}r@{\hspace{0.2cm}}r@{\hspace{0.2cm}}r@{\hspace{0.2cm}}r@{\hspace{0.2cm}}r}
\multicolumn{2}{l}{Graphs}      &d      &$\mu$       &tomita &maxdegree&hybrid &degen \\ \cline{1-2}
n                      &p       &       &            &       &         &       &      \\ \hline
\multirow{4}{*}{100}   &0.6     &51     &59,898      &0.08   &0.26     &0.25   &0.14  \\
                       &0.7     &59     &439,928     &0.50   &2.04     &1.85   &0.99  \\
                       &0.8     &70     &5,776,276   &6.29   &28.00    &24.86  &11.74 \\
                       &0.9     &81     &240,998,654 &249.15 &1136.15  &1028.84&425.85\\ \\
\multirow{6}{*}{300}   &0.1     &21     &3,663       &$<0.01$&0.01     &0.01   &$<0.01$  \\
                       &0.2     &47     &18,911      &0.02   &0.07     &0.08   &0.05  \\
                       &0.3     &74     &86,179      &0.10   &0.44     &0.49   &0.24  \\
                       &0.4     &101    &555,724     &0.70   &4.24     &3.97   &1.67  \\
                       &0.5     &130    &4,151,668   &5.59   &42.37    &36.35  &13.05 \\
                       &0.6     &162    &72,454,791  &101.35 &958.74   &755.86 &227.00\\ \\
\multirow{4}{*}{500}   &0.1     &39     &15,311      &0.02   &0.03     &0.06   &0.04  \\
                       &0.2     &81     &98,875      &0.11   &0.46     &0.61   &0.27  \\
                       &0.3     &127    &701,292     &0.86   &5.90     &6.10   &2.29  \\
                       &0.5     &225    &103,686,974 &151.67 &1888.20  &1521.90&375.23\\ \\
\multirow{4}{*}{700}   &0.1     &56     &38,139      &0.04   &0.10     &0.19   &0.09  \\
                       &0.2     &117    &321,245     &0.37   &2.01     &2.69   &1.00  \\
                       &0.3     &184    &3,107,208   &4.06   &36.13    &38.12  &11.47 \\ \\
\multirow{3}{*}{1,000} &0.1     &82     &99,561      &0.11   &0.34     &0.70   &0.28  \\
                       &0.2     &172    &1,190,899   &1.45   &10.35    &14.48  &4.33  \\
                       &0.3     &266    &15,671,489  &21.96  &262.64   &280.58 &66.05 \\ \\
2,000                  &0.1     &170    &750,991     &1.05   &5.18     &11.77  &3.13  \\ \\
3,000                  &0.1     &263    &2,886,628   &4.23   &27.51    &68.52  &13.62 \\ \\
\multirow{5}{*}{10,000}&0.001   &7      &49,716      &1.19   &0.04     &0.07   &0.07  \\
                       &0.003   &21     &141,865     &1.30   &0.11     &0.36   &0.26  \\
                       &0.005   &38     &215,477     &1.47   &0.25     &1.03   &0.51  \\
                       &0.01    &80     &349,244     &2.20   &1.01     &5.71   &1.66  \\
                       &0.03    &262    &3,733,699   &9.96   &20.66    &133.94 &20.67
\end{tabular}
\end{center}
\end{table}

\section{Conclusion} We have shown that the algorithm of Eppstein, L\"offler, 
and Strash is a practical algorithm for large sparse graphs. This algorithm
is highly competitive with the algorithm of Tomita et al. on sparse graphs,
and within a small constant factor on other graphs. The advantage of this
algorithm is that it requires only linear space for storing the graph
and all data structures. It does not suffer from the drawback of 
requiring an adjacency matrix, which may not fit into memory.  Its closest competitor in this respect, the Tomita et al. algorithm modified to use adjacency lists, is sometimes faster by a small factor but is also sometimes slower by a large factor. Thus, the algorithm of Eppstein et al. is a fast and reliable choice for listing maximal cliques, especially when the input graphs are large and sparse. 

For future work, it would be interesting to compare our results with those of other popular clique listing algorithms. We attempted to include results from Patric \"Osterg\aa rd's popular Cliquer program~\cite{cliquer} in our tables; however, at the time of writing, its newly implemented functionality for listing all maximal cliques returns incorrect results.

\subsubsection*{Acknowledgments} We thank Etsuji Tomita and Takeaki Uno for helpful discussions.
This research was supported in part by the National Science Foundation under grant
0830403, and by the Office of Naval Research under MURI grant N00014-08-1-1015.


\end{document}